\documentclass[11pt]{article}

\usepackage[DIV13]{typearea}
\usepackage{amsmath}
\usepackage{amsfonts}

\usepackage{amssymb}
\usepackage[titletoc]{appendix}
\usepackage{array}
\usepackage{bbold} 
\usepackage{bbm}
\usepackage{booktabs}
 \usepackage{cancel}
 \usepackage{subcaption}
 \usepackage{cite} 
\usepackage[usenames,dvipsnames]{color}
\usepackage{epsfig}
\usepackage{fancyhdr}
\usepackage[T1]{fontenc}
\usepackage{framed}
\usepackage[left=.9in, right=.9in]{geometry}
\usepackage{graphicx}
\usepackage{hhline} 
\RequirePackage[colorlinks=true,urlcolor=blue,anchorcolor=blue,citecolor=blue,filecolor=blue,
               linkcolor=blue,menucolor=blue,linktocpage=true,pdfproducer=medialab,pagebackref=true]{hyperref}
\usepackage[utf8]{inputenc}
\usepackage{mathrsfs}
\usepackage{mathtools}
\usepackage{dsfont}
\usepackage{multicol}
\usepackage{multirow}
\usepackage{indentfirst}
\usepackage{longtable}
 \usepackage{lscape}
\usepackage{pbox}
\usepackage{pdfpages}
\usepackage{rotating}
\usepackage{slashed}
\usepackage[DIV13]{typearea}
\usepackage[normalem]{ulem}
\usepackage{units}
\usepackage{enumitem}
\usepackage{soul}
\usepackage{placeins}
\usepackage{tikz}

\hypersetup{
  colorlinks   = true, 
  urlcolor     = magenta, 
  linkcolor    = blue, 
  citecolor   = blue 
}

\numberwithin{equation}{section}

\usepackage{csquotes}

\newcommand{\blue}[1]{\color{blue}#1\color{black}}

\newcommand{\beq}{\begin{equation}}%
\newcommand{\eeq}{\end{equation}}%
\newcommand{\bea}{\begin{align}}%
\newcommand{\ena}{\end{align}}%

\usepackage{catchfile}
\newcommand{\getenv}[2][]{%
  \CatchFileEdef{\temp}{"|kpsewhich --var-value #2"}{}%
  \if\relax\detokenize{#1}\relax\temp\else\let#1\temp\fi
}
\getenv[\USER]{USER}

\usepackage{catchfile}

\allowdisplaybreaks
\renewcommand{\to}{\rightarrow}

\renewcommand{\[}{\begin{equation}}
\renewcommand{\]}{\end{equation}}
\definecolor{orange}{rgb}{1,0.5,0}

\newcounter{diagram}

\newcommand{\email}[1]{\href{mailto:#1}{\tt #1}}

\renewcommand{\to}{\rightarrow}
\newcommand{\nn}{\nonumber}
\renewcommand{\to}{\longrightarrow}

\newcommand{\fpq}{f_{\text{PQ}}}

\definecolor{green}{rgb}{0.13, 0.55, 0.13}

\newcolumntype{C}[1]{>{\centering\let\newline\\\arraybackslash\hspace{0pt}}m{#1}}

\begin{document}
\vskip 1.5cm
\vspace*{-1cm}
\phantom{hep-ph/***}

{\flushright
{\blue{IPMU18-0205}\\
\blue{IFT-UAM/CSIC-18-129}\\
\blue{FTUAM-18-29}}\\
\hfill 
}
\vskip 1.5cm
\begin{center}
{\LARGE  Automatic Peccei-Quinn symmetry\\[0.5cm] }
\vskip .3cm
\end{center}
\vskip 0.5  cm
\begin{center}

{\large M.B.~Gavela}~$^{a)}$,
{\large M.~Ibe}~$^{b,c)}$, 
{\large P.~Quilez}~$^{a)}$,
{\large T. T.~Yanagida}~$^{c,d)}$
\\
\vskip .7cm
{\footnotesize
$^{a)}$~
Departamento de F\'isica Te\'orica and Instituto de F\'{\i}sica Te\'orica, IFT-UAM/CSIC,\\
Universidad Aut\'onoma de Madrid, Cantoblanco, 28049, Madrid, Spain\\
\vskip .1cm
$^{b)}$Institute for Cosmic Ray Research, University of Tokyo, Kashiwa, Chiba 277-8582, Japan\\
\vskip .1cm
$^{c)}$Kavli Institute for the Physics and Mathematics of the Universe (WPI), \\University of Tokyo, Kashiwa, Chiba 277-8583, Japan\\
\vskip .1cm
$^{d)}$T. D. Lee Institute and School of Physics and Astronomy,\\
Shanghai Jiao Tong University, Shanghai 200240, China
\vskip .5cm
}
\end{center}

\begin{abstract}

We present a dynamical (composite) axion model where the Peccei-Quinn (PQ) symmetry arises automatically as a consequence of chirality and gauge symmetry. The Standard Model is simply extended by a confining and chiral $SU(5)$ gauge symmetry. The PQ symmetry coincides with a $B-L$ symmetry of the exotic sector. The theory is protected by construction from quantum gravitational corrections stemming from operators with mass dimension lower than nine.

\end{abstract}

\vskip 1cm

\begin{minipage}[l]{.9\textwidth}
{\footnotesize
\begin{center}
\textit{E-mail:} 
\email{belen.gavela@uam.es}\,,
\email{ibe@icrr.u-tokyo.ac.jp}\,,
\email{pablo.quilez@uam.es}\,,
\email{tsutomu.tyanagida@ipmu.jp}
\end{center}}
\end{minipage}

\pagebreak
\tableofcontents

\pagebreak

%
%
\section{Introduction} \label{introduction}
A most intriguing puzzle of the Standard Model of Particle Physics (SM) is the so-called ``strong CP problem'': the  extremely small size  ($< 10^{-10}$~\cite{Baker:2006ts,Engel:2013lsa}) 
of the vacuum angle of the strong interactions
\begin{equation}
  |\bar{\theta}|\,=\, |\theta_{QCD} + \text{arg det}\,M|\,.
  \end{equation}
Here, $M$ denotes the quark mass matrix and $\theta_{QCD}$ characterises the CP-odd gauge contribution in the $SU(3)_c$ QCD Lagrangian, 
  \begin{equation}
  \mathcal{L} \,=\, -\frac{1}{4} G_{\mu \nu}G^{\mu \nu} \,-\, \theta_{QCD}\,\frac{\alpha_s}{8\pi} G_{\mu \nu}\tilde{G}^{\mu \nu} \,+\, \bar{q}\,M\,q\,,
  \end{equation}
where  $\alpha_s$ denotes the QCD fine structure constant and colour indices have been left implicit. 

The most elegant solution to the strong CP problem is to introducte a global chiral $U(1)$ symmetry, usually called Peccei-Quinn (PQ) symmetry~\cite{Peccei:1977hh} $U(1)_{PQ}$, which is exact (and hidden) at the classical level but is anomalous under QCD interactions. The latter is the key to solve the problem and  also the only source of  
the mass for the pseudo-Nambu-Goldstone boson of the global $U(1)_{PQ}$ symmetry: the axion.

A simple and most economical implementation  would be the  $U(1)_{PQ}$  symmetry that would exist if one SM quark were to be massless. 
The freedom to chirally rotate  that fermion would allow to fully reabsorb all contributions to $\bar{\theta}$, making it unphysical. 
This interesting possibility~\cite{Georgi:1981be} does not seem to be realized in nature after the  constraints stemming from lattice computations and we disregard it, even if the option is not completely excluded~\cite{Dine:2014dga, Choi:1988sy, Kaplan:1986ru, Frison:2016rnq, Frison:2017mod,Aoki:2016frl,Bardeen:2018fej}.

It is still possible to solve the strong CP problem with massless quarks, though, if extra exotic massless fermions charged under QCD  exist in Nature. 
 As the latter are not observed, the idea~\cite{Choi:1985cb} is to charge them in addition under  a new confining force~\cite{Weinberg:1975gm,Susskind:1978ms,Dimopoulos:1981xc}, often called ``axicolor''~\cite{Choi:1985cb}, whose scale is much larger than that of QCD, $\Lambda_{QCD}$.  A new spectrum of confined states results composed of those massless fermions, including mesons which play the role of axions. 
 They  are often referred to as ``dynamical'' or composite axions.

 In a given theory, when the number of axions --either elementary or composite--  outnumbers the total number of distinct instanton-induced scales other than QCD, 
 one (or more) light axions remain.    
 These are called ``invisible axions'', whose mass $m_a$ and scale $f_a$  generically obey~\cite{Weinberg:1977ma,Georgi:1986df} 
\begin{equation}
 m_a^2 f_a^2 \sim m_\pi^2 f_\pi^2\,\frac{m_u\,m_d}{(m_u+m_d)^2}\,,
 \label{invisiblesaxion}
 \end{equation}
where 
 $m_\pi, f_\pi, m_u, m_d$ denote 
 the pion mass and coupling constant, and the up  and down quark masses, respectively.%
 \footnote{Alternative models with extra sources of instantons may render all axions heavier than the QCD scale, and they are increasingly explored in the last 
 years~\cite{Rubakov:1997vp,Berezhiani:2000gh,Hsu:2004mf,Hook:2014cda,Fukuda:2015ana,Chiang:2016eav,Dimopoulos:2016lvn,Gherghetta:2016fhp,Kobakhidze:2016rwh,Agrawal:2017ksf,Agrawal:2017evu,Gaillard:2018xgk}.}   Light enough axions {(that is, below $\mathcal{O}(100\,\mathrm{MeV})$)} can participate in astrophysical phenomena~\cite{Raffelt:2006cw,Chang:2018rso,Irastorza:2018dyq,Hamaguchi:2018oqw}.  
 The constraints that follow from their non-observation in photonic processes lead to  very high values for {the decay constant},  $f_a\ge10^8$\,GeV. 
 It follows then from Eq.\,(\ref{invisiblesaxion}) that  $m_a \le 10^{-2}$\,eV.
 Here we will construct a novel implementation of the invisible axion paradigm via massless exotic quarks, and in consequence Eq.\,(\ref{invisiblesaxion}) will apply.

In the original composite axion proposal~\cite{Choi:1985cb} the confining sector of the SM was enlarged to $SU(3)_c \times SU(\tilde{N})$, where  $SU(\tilde{N})$ is the axicolor group.
Two composite axions result, one of which must be invisible and obey Eq.\,(\ref{invisiblesaxion}), as there are only two sources of instantons for three pseudoscalars with anomalous couplings (taking into account the SM $\eta'$).
The axicolor construction can be seen as a beautiful ultraviolet dynamical completion of the invisible axion paradigm. 
It has the advantage of being free from the scalar potential fine-tunings that hinder models of invisible elementary axions {\it \`a la} KSVZ~\cite{Kim:1979if,Shifman:1979if} or DFSZ~\cite{Zhitnitsky:1980tq,Dine:1981rt}. 
 
Dynamical axion constructions often require that the PQ transition predates inflation. 
This avoids cosmological problems in the form of domain walls (whose accumulated energy could  overclose the universe  after the QCD phase transition).   
Our patch of the universe would correspond  to a specific initial value of the axion field which determines the axion energy density, because of the misalignment mechanism~\cite{Turner:1985si}. 
In the absence of fine-tuned values of the misalignment angle, if axions were to explain all the dark matter density  it is necessary that~\cite{Visinelli:2009zm,Saikawa17p083} 
\begin{equation}
f_a\simeq 2\times10^{10}- 5\times10^{12}\,\text{GeV}\,,  
\label{dmrange}
\end{equation} 
although the axion decay constant could be one order of magnitude smaller if some fine tuning is allowed.
 
  A threat which menaces  all types of invisible axion models stems from quantum  non-perturbative gravitational corrections~\cite{Holman:1992us,Kamionkowski:1992mf, Barr:1992qq, Ghigna:1992iv, Georgi:1981pu, Giddings:1988cx,Coleman:1988tj,Gilbert:1989nq, Rey:1989mg}, as  $f_a$  is not very far from the Planck scale.  
These are usually parametrized via effective operators  suppressed by powers of the Planck mass, $M_{\text{\rm Pl}}$.%
\footnote{Here, the Planck mass does not denote the reduced Planck scale but the one given by $M_{\rm Pl} = G^{-1/2}$ with $G$ being the Newton constant.}
They would  explicitly violate the PQ symmetry and  can thus spoil the solution to the SM strong CP problem. For instance, Ref.~\cite{Holman:1992us,Kamionkowski:1992mf, Barr:1992qq, Ghigna:1992iv} concentrated on the simplest (and most dangerous) hypothetical dimension five effective operator 
\begin{equation}
g_5\, \frac {|\Phi|^4\, (\Phi+ \Phi^*) }{M_{\rm{\rm Pl}}} \,,
\label{gravity-5}
\end{equation}
where $g_5$ is a dimensionless coefficient and $\Phi$ would be a field whose VEV breaks the PQ invariance. 
In order to avoid that this term moves the minimum of the axion potential unacceptably away from  a CP-conserving solution, its coefficient needs a extreme fine-tuning, e.g. $g_5< 10^{-54}$ for $f_a\sim 10^{12}$\,GeV.%
\footnote{They can be avoided, though, in some invisible axion constructions with a variety of extra assumptions or 
frameworks~\cite{Randall:1992ut,Dobrescu:1996jp,Butter:2005wr,Redi:2016esr, Fukuda:2017ylt,Fukuda:2018oco,Ibe:2018hir,Lillard:2018fdt}, 
or be arguably negligible in certain conditions~\cite{Alonso:2017avz}. 
It is also possible to avoid the dangerous terms in ``heavy axion'' models~\cite{Rubakov:1997vp,Fukuda:2015ana,Berezhiani:2000gh,Hsu:2004mf,Hook:2014cda,Chiang:2016eav,Dimopoulos:2016lvn,Gherghetta:2016fhp,Kobakhidze:2016rwh,Agrawal:2017ksf,Agrawal:2017evu,Gaillard:2018xgk}, as their $f_a$ scale can be very low, e.g. not far from the TeV range.
}  
 
 In this work, the axicolor framework is approached with a novel light: to assume that the  $SU(\tilde{N})$ exotic confining gauge sector is chiral.  
 In a minimalistic approach, we require   a fermion content such that:
 \begin{itemize}
 \item It confines and renders the  theory  free from gauge anomalies. 
 \item The exotic fermion representations are chiral,  so that fermionic mass terms are automatically forbidden.
 \item  Minimality in the specific matter content will be a guideline. Two (or more) different axicolored fermions are present, with at least one of them being QCD colored as well. 
 \end{itemize}
In this class of set up, at least two chiral $U(1)$ symmetries emerge in the dynamical sector in the limit of $M_{\rm Pl} \to \infty$ and nullify the theta angles of the 
dynamical sector and the QCD sector.
It can be checked that it is not possible to obey the three requirements listed above for $SU(3)$, $SU(6)$ or $SU(7)$, at least not with just two exotic fermions in low-dimensional representations of the chiral confining group.
It is possible instead for $SU(4)$; nevertheless, this theory would not render an improvement on the gravitational issue, as argued in App.~\ref{App:SU(4)}, and it will not be further developed.
 
 We focus here on the case of chiral gauge $SU(5)$, implemented via its lowest dimensional 
 fermion representations, $\mathbf{\bar{5}}$ and  $\mathbf{10}$, which together fulfil the conditions above.
The $SU(5)$ confinement scale will be  assumed to  be much larger than that of QCD, $\Lambda_5 \gg\Lambda_{\rm QCD}$. 
It will be shown that a satisfactory $U(1)_{PQ}$ symmetry is an automatic consequence of the chiral realization of the gauge group. 
Note that some models have been previously built  for which PQ invariance is  accidental, that is, not imposed by hand ~\cite{Randall:1992ut,Dobrescu:1996jp,Butter:2005wr,Redi:2016esr, Fukuda:2017ylt,Fukuda:2018oco,Ibe:2018hir,Lillard:2018fdt}.
Nevertheless,  they all required extra symmetries in addition to axicolor, either gauge or discrete ones. 
In contrast,  axicolor $SU(5)$  will be shown to suffice because of its chiral character, rendering a particularly simple framework.

Relevant aspects to be developed include on one side the identification of the exotic fermion condensates, which in dynamical axion models are  the only source of spontaneous symmetry breaking, e.g. for exotic flavour and  for the PQ symmetries.  Another important question is the impact of $SU(5)$ gauge invariance on the possible  non-perturbative gravitational couplings of the theory.

The idea will be implemented in two alternative realizations, selected so as to achieve minimal matter content. 
They will only differ in the QCD charges of the exotic $\mathbf{\bar{5}}$ and  $\mathbf{10}$ fermions present:  
octets of QCD color in one model, while triplets in a second version. 
 
The structure of the paper can be easily inferred from the Table of Contents.

\section{The SU(5) chiral confining theory}\label{sec:model}
We consider  a chiral version of the axicolor model, with $SU(5)$ as an extra confining group,  and one set  of massless exotic fermions in its five  and ten dimensional representations, $\psi_{\bar{5}}$ and $\psi_{10}$ (the notation  $\psi_{\bar{5}}\equiv \mathbf{\bar{5}},\, \psi_{10}\equiv \mathbf{10}$ will be often used for convenience). Such a set cancels   
 all $SU(5)$ gauge anomalies (as in $SU(5)$ GUT models).   The complete gauge group of Nature would then be 
   \begin{equation}
  SU(5)\times SU(3)_c \times SU(2)_L\times U(1)\, .
  \end{equation}
  An economic implementation is to assume the usual SM fields  to be  singlets under $SU(5)$,  while the exotic chiral fermions   in the  $\psi_{\bar{5}}$ and $\psi_{10}$ representations  of $SU(5)$ are singlets under the electroweak SM gauge group.
 \begin{table}[h!]
\begin{align*}
\begin{array}{c|c|c|}
	& SU(5)	& SU(3)_c	\\
\hline
\psi_{\bar{5}}	& \mathbf{\bar{5}}		& \mathbf{R}	\\
\psi_{10}\,	& \mathbf{10}		& \mathbf{R}	
\end{array}
\end{align*} 
\caption{Charges of exotic fermions under the confining gauge group $SU(5)\times SU(3)_c$.  The left-handed Weyl fermions $\psi_{\bar{5}}$ and $\psi_{10}$ are massless and singlets of the SM electroweak gauge group. $\mathbf{R}$ denotes a pseudoreal representation. }
\label{tab:matterContent}
\end{table}

  If the exotic fermions carry also QCD color, this theory solves the strong CP problem. Indeed, the presence of   (at least) two massless fermions  ensures the existence of two distinct $U(1)$ chiral global symmetries, exact at the classical level but explicitly broken by quantum non-perturbative effects. The
$\theta$-parameters corresponding to the two confining gauge groups become thus unphysical via chiral rotations of those fermions. 
  Furthermore, the chiral character of the representations forbids fermionic mass terms and thus guarantees that those symmetries are automatic, instead of imposed on a given Lagrangian as customary.    Finally,  the requirement of a large confining scale $\Lambda_5\gg \Lambda_{\rm QCD}$  leads to a realistic model, given the non-observation of a spectrum of bound states composed of those massless exotic fermions.

For simplicity, we will consider that the set $\{\psi_{\bar{5}}, \psi_{10}\}$ belongs to a (pseudo)real representation  $\mathbf{R}$ of color QCD, so as automatically cancel $\left[SU(3)_c\right]^3$ anomalies, see Table~\ref{tab:matterContent}. 
Later on we will develop in detail two specific choices for $\mathbf{R}$:  the case of the fundamental of QCD with reducible representation $\mathbf{R}=  \mathbf{3}+ \mathbf{\bar{3}}$ in one case, and  the adjoint $\mathbf{R}=\mathbf{8}$ in the second case. In all cases, all mixed gauge anomalies in the confining sector vanish by construction as well, because only non-abelian $SU(N)$ groups are present and the exotic fermions are electroweak singlets.

\subsection{Global symmetries}

At the scale $\Lambda_5$,  $SU(5)$ confines and  the massless fermions in Tab.~\ref{tab:matterContent} will form massive bound states including QCD-colored ones. In the limit in which the QCD coupling constant $\alpha_s$ is neglected, the $SU(5)$ gauge Lagrangian exhibits at the classical level a global flavor symmetry
\begin{equation}
 U(n)_{\bar{5}} \times U(n)_{10} \leftrightarrow  SU(n)_{\bar{5}} \times SU(n)_{10}\times U(1)_{\bar{5}}\times U(1)_{10}\,,
 \label{globalsym}
 \end{equation}
 where $n$ denotes the dimension of $\mathbf{R}$, which plays the role of number of exotic flavours, 
 \begin{equation}
n= \text{dim} \{\mathbf{R}\}\,.
 \end{equation}
The  two global $U(1)$ symmetries correspond to independent rotations of the two massless fermion representations. However, they are both broken at the quantum level by anomalous couplings to the $SU(5)$ and QCD field strengths. 
 A generic combination of them 
will lead to the following anomaly coefficients (see App.~\ref{App:Anomaly}): 
\begin{align}
U(1)\times \left[SU(5)\right]^2:&\qquad n\times \big(Q_{\bar{5}}T(\mathbf{\bar{5}})+Q_{10}T(\mathbf{10})\big)=\frac{n}{2}\left(Q_{\bar{5}}+3\, Q_{10}\right)\,, \label{comb1}\\
U(1)\times \left[SU(3)_c\right]^2:&\qquad  T(\text{\textbf{R}})\times\big(5\, Q_{\bar{5}}+10\, Q_{10}\big) \label{comb2}\,.
\end{align}
Here, $Q_{\bar{5}}$ and $Q_{10}$ denote arbitrary $U(1)$ charges for $\psi_{\bar{5}}$ and  $\psi_{10}$, respectively,
and $T$'s denote the  Dynkin indices of the corresponding representations.
It follows from Eq.\,(\ref{comb1}) that the charge assignment \begin{equation}
Q_{\bar{5}}=-3\,,\qquad Q_{10}=1\,,
\end{equation} 
renders a combination of $U(1)$'s that is free from $SU(5)$ anomaly.  
The $SU(5)$ anomaly-free combination is analogous to the $B-L$ symmetry in usual $SU(5)$ GUT's. 
It will play the role of the PQ symmetry in our model, since it is a classically exact symmetry that is only broken by the QCD anomaly.   
A second combination will remain explicitly broken%
\footnote{This can be for instance, the orthogonal combination corresponding to $\{Q_{\bar{5}}=1,\,Q_{10}=3\}$, although any  combination different from that free from anomalous $SU(5)$ couplings can play this role.} by quantum non-perturbative effects of $SU(5)$, 
so that the classical global symmetry in Eq.\,(\ref{globalsym}) reduces (for $\alpha_s=0$)  to
\begin{equation}
SU(n)_{\bar{5}} \times SU(n)_{10}\times U(1)_{PQ=B-L} \,.
\label{chiralsym}
\end{equation}
The  corresponding global charges of the exotic fermions  are shown in Table~\ref{tab:GlobalSym}. 
\begin{table}[h!]
\begin{align*}
\begin{array}{c|c|c|c|c}
	& SU(n)_{\bar{5}}	& SU(n)_{10}	& U(1)_{B-L}\equiv U(1)_{PQ} 		\\
\hline
\psi_{\bar{5}}	& \Box	& \mathbf{1}	& -3		\\
\psi_{10}\,	& \mathbf{1}		& \Box & 1			
\end{array}
\end{align*} 
\caption{Global chiral properties at the classical level, in the limit of vanishing $\alpha_s$. }
\label{tab:GlobalSym}
\end{table}

\subsubsection*{Confinement versus chiral symmetry breaking}
A first question is whether the confinement of the $SU(5)$ gauge dynamics is accompanied by the spontaneous breaking of  the associated chiral global symmetries. Two alternative realizations are possible:
\begin {itemize}
\item The global symmetries can be spontaneously broken via fermion condensates. As a result,  (almost) massless (pseudo)Goldstone bosons (pGBs) will be present in the low energy theory. 
\item Conversely, they could remain unbroken and  the spectrum of bound states would explicitly reflect those global symmetries via multiplets of degenerate states. In particular, massless baryons are then needed in order to fulfil the `t Hooft anomaly consistency conditions~\cite{tHooft:1979rat} to match the anomalies of the high and low energy theories.
\end{itemize}
It can be shown that it is not possible to comply with the 't Hooft consistency conditions for the complete flavour group. That is, it is impossible to match the $\left[SU(n)_{\bar{5}}\right]^3$ and $\left[SU(n)_{10}\right]^3$ anomalies before confinement --and thus in terms of quarks-- with the anomalies after confinement in terms of  massless ``baryons''. 
The demonstration can be found in App.~\ref{App:tHooft}.   
The confinement of gauge $SU(5)$ is thus necessarily accompanied by the spontaneous breaking of the chiral global $SU(n)_{\bar{5}}\times SU(n)_{10}$ symmetry, 
and  associated (pseudo)Goldstone bosons (pGBs) will be present in the low-energy spectrum.

In contrast, for $U(1)_{PQ}$ it is possible 
to fulfil 't Hooft anomaly conditions~\cite{Dimopoulos:1980hn,ArkaniHamed:1998pf}. 
 At high energies and in terms of quarks, the spectrum in Tab.~\ref{tab:GlobalSym}  contributes to the global anomalies as
 \begin{align}
\left[U(1)_{PQ}\right]^3:& \qquad n \left(5\,(Q_{\bar{5}})^3+10\,(Q_{10})^ 3\right)=-125 \,n\,, \label{PQ3}\\
U(1)_{PQ}\times \left[SU(3)_c\right]^2:& \qquad N\equiv 2\,(5\,Q_{\bar{5}}  T(\mathbf{R})+10\,Q_{10} T(\mathbf{R}))=-10\,T(\mathbf{R}) \,, \\
U(1)_{PQ}\times \left[\text{grav}\right]^2:& \qquad n\,(5\,Q_{\bar{5}}+10\,Q_{10})=-5\,n \,.
\label{Eq: PQ QCD anomaly}
\end{align}
where $N$ denotes as customary the QCD anomaly factor. The low-energy spectrum admits in turn a  massless baryon composed by three fermions, 
\begin{equation} 
\chi\equiv\mathbf{10}\, \mathbf{\bar{5}} \, \mathbf{\bar{5}}\,,
\label{onlybaryon}
\end{equation}
which has  PQ charge $Q_\chi=-5$ and can belong to the $\mathbf{R}$ representation of $SU(3)_c$.
  Its contribution to the anomaly equations matches  the anomalies at the quark level in Eqs.\,(\ref{PQ3}) and (\ref{Eq: PQ QCD anomaly}): 
\begin{align}
\left[U(1)_{PQ}\right]^3:& \qquad n \,Q_\chi^3=-125\,n\,,\\
U(1)_{PQ}\times \left[SU(3)_c\right]^2:& \qquad N\equiv 2\,Q_\chi\, T(\mathbf{R})=-10\,T(\mathbf{R})\,,\\
U(1)_{PQ}\times \left[\text{grav}\right]^2:& \qquad n\,Q_{\chi}=-5\,n \,, 
\end{align}
In consequence, the chiral confining $SU(5)$ theory  would be {\it a priori } perfectly consistent even if the $U(1)_{PQ}$ were to remain unbroken after confinement. Nevertheless,  this is not phenomenologically viable since  (almost) massless  QCD colored fermions are not  observed in Nature (other than the light SM quarks). 
 
 To sum up,  parts of the global symmetries in Eq.\,(\ref{chiralsym}) with the field content in Table~\ref{tab:GlobalSym} need to be spontaneously broken by fermion condensates upon $SU(5)$ confinement.

\subsection{Fermion condensates:  chiral-breaking versus PQ-breaking}

It will be assumed that $\Lambda_5$ settles the overall scale for all dynamical breaking mechanisms in the $SU(5)$ sector, which will take place through fermion condensates.


\subsubsection*{Chiral condensate}  
The lowest dimension fermionic condensate which is gauge invariant and breaks the non-abelian chiral symmetries in Eq.\,(\ref{chiralsym}) is a dimension six operator: 
\begin{equation}
\mathbf{10\,10\,10\,\bar{5}}\,,
\end{equation}
 with vacuum expectation value (VEV)  and breaking pattern expected to obey   
\begin{equation}
\langle \mathbf{10\,10\,10\,\bar{5}}\rangle \sim \Lambda_5^6 \quad \implies \quad SU(n)_{\bar{5}} \times SU(n)_{10}\longrightarrow G\supset SU(3)_c \,.
\label{Eq:chiral condensate SU(n)}
\end{equation}
On the right-hand side of this expression, it has been assumed that the QCD gauge group is contained in 
the unbroken subgroup $G$ of $ SU(n)_{\bar{5}} \times SU(n)_{10}$.
This is possible as the product of four $\mathbf{R}$ representations contains an $SU(3)$ singlet since $\mathbf{R}$ is (pseudo)real. 
It should be noted that the unbroken subgroup $G$ which contains $SU(3)$ is not necessarily aligned 
with the one which contains $SU(3)_c$ for $\alpha_s = 0$.
Once  $\alpha_s$  is turned on, on the other hand, the QCD interaction forces the condensates to preserve color, 
which implies that only the QCD invariant condensates will form (see also\cite{Dobrescu:1996jp}).%
\footnote{In the thermal bath, for example, the QCD breaking vacua have higher energy density
than the QCD preserving one due to the thermal potential proportional proportional to $m_{\rm gluon}^2T^2$,  
where $m_{\rm gluon}$ denotes the gluon mass on the QCD breaking vacua.
}

If \textbf{R} is an irreducible representation of $SU(3)_c$, then  the only part of the non-abelian chiral symmetry in Eq.\,(\ref{Eq:chiral condensate SU(n)}) that is expected to remain unbroken is $SU(3)_c$. If  \textbf{R} is reducible instead, some $U(1)$'s can remain exact  (see Sec.~\ref{Sec:3+3} where $\mathbf{R}=\mathbf{3}+\mathbf{\bar{3}}$). Therefore, irrespective of $G$,  most generators of $SU(n)_{\bar{5}} \times SU(n)_{10}$ other than those of $SU(3)_c$   would be explicitly broken by QCD interactions,
\begin{equation}
 SU(n)_{\bar{5}} \times SU(n)_{10} \xrightarrow[]{\langle \mathbf{10\,10\,10\,\bar{5}}\rangle}G \xrightarrow{\alpha_s \neq 0}SU(3)_c \,. 
 \label{chiralbreaking}
\end{equation}
In consequence, most of the pGBs associated to the broken generators of the non-abelian chiral symmetry are necessarily colored under QCD.  
Their masses $m$ are quadratically sensitive to large scales via gluon loops and thus safely large, 

  \begin{equation}
m^2(\text{{\bf R}})\sim \frac{3\,\alpha_s}{4\pi}\,C(\text{{\bf R}})\,\Lambda_5^2\,,
\end{equation}
where C(\text{{\bf R}}) is the quadratic Casimir of the QCD representation {\bf R} to which a given pGB belongs, $T^a_{R} T^a_{R}=C(\text{{\bf R}}) \,\mathbb{1} $.
 
The chiral condensate in Eq.\,(\ref{Eq:chiral condensate SU(n)}) is $U(1)_{PQ}$ invariant, though, since its PQ charge is vanishing. The spontaneous breaking of the PQ symmetry (which is phenomenologically the only viable option as earlier explained) can only be achieved via higher dimensional fermionic condensates.

\subsubsection*{PQ condensate}
The lowest dimensional operator  which is gauge invariant  but has non-vanishing PQ-charge is 
\begin{equation}
 \mathbf{\bar{5}\,\bar{5}\,10\,\bar{5}\,\bar{5}\,10} \, ,
 \label{Eq:PQ condensate SU(n)}
\end{equation}
which has mass dimension nine and PQ-charge -10. In order to achieve spontaneous $U(1)_{PQ}$ symmetry breaking, we assume that this operator obtains a non-vanishing VEV,\footnote{Its VEV also breaks the non-abelian chiral symmetry, but this effect should be subdominant with respect to that of the lower dimension operator in Eq.\,(\ref{Eq:chiral condensate SU(n)}).} 
\begin{equation}
 \langle \mathbf{\bar{5}\,\bar{5}\,10\,\bar{5}\,\bar{5}\,10}\rangle \sim \Lambda_5^9\,, 
 \label{Eq:PQ condensate SU(n)}
\end{equation}
which is associated with the QCD axion as a composite field.

In summary, the combined action of the two condensates in Eqs.\,(\ref{Eq:chiral condensate SU(n)})  and (\ref{Eq:PQ condensate SU(n)}) induces a breaking pattern of the global symmetries of the exotic $SU(5)$ sector of the form 
\begin{eqnarray}
SU(n)_{\bar{5}} \times SU(n)_{10}\times U(1)_{PQ} 
 \xrightarrow{\langle \mathbf{10\,10\,10\,\bar{5}}\rangle}G\times U(1)_{PQ}  
 \xrightarrow{\langle \mathbf{\bar{5}\,\bar{5}\,10\,\bar{5}\,\bar{5}\,10}\rangle} G' 
 \xrightarrow{\alpha_s \neq 0}SU(3)_c\,.  
\label{totalbreaking}
\end{eqnarray}
For later use, it is convenient to parametrize the field combination in Eq.\,(\ref{Eq:PQ condensate SU(n)})  as
\begin{equation}
 \mathbf{\bar{5}\,\bar{5}\,10\,\bar{5}\,\bar{5}\,10} \sim \Lambda_5^9\,e^{-i\,10 \, a/\fpq}\,, \label{Eq:PQ condensate SU(8)}
\end{equation}
 
where the radial degrees of freedom are left implicit,  $a$ denotes the dynamical axion that corresponds to the axial excitation of the operator, and 
 the PQ charge of the condensate resulting from Tab.~\ref{tab:GlobalSym} is explicitly shown. The PQ scale $f_{PQ}$ associated to the pGB nature of the axion 
 obeys 
\begin{equation}
\fpq \propto \Lambda_5\,.
\end{equation}
It should be noted that the PQ charges of the $SU(5)$ invariant states are multiples of $5$, and hence,
the PQ symmetry in the broken phase is realized by a shift of the axion given by 
\begin{eqnarray}
\label{eq:shift}
\frac{5 a}{f_{PQ}} \to \frac{5 a}{f_{PQ}} + \alpha\ , \quad \alpha = [0,2\pi) \ ,
\end{eqnarray}
see also App.\,\ref{App:Axion domain}.

\subsection{The axion Lagrangian}
 In order to  obtain the low-energy effective Lagrangian for the axion,   the conservation of the PQ current will be studied next. The current at high energies can be computed in terms of the fundamental fermions by applying Noether's formula: 
\begin{equation}
j^\mu_{\text{PQ}}=Q_5 \,\psi_{\bar{5}}^\dagger \bar \sigma ^\mu \psi_{\bar{5}} + Q_{10}\,
\psi_{10}^\dagger \bar \sigma ^\mu \psi_{10}=-3 \,\psi_{\bar{5}}^\dagger \bar \sigma ^\mu \psi_{\bar{5}} + 
\psi_{10}^\dagger \bar \sigma ^\mu \psi_{10}=f_{\text{PQ}}\partial^\mu a\,.
\label{PQcurrent}
\end{equation}
At energies below $SU(5)$ confinement, the current can be expressed in terms of the composite fermions (i.e. the composite baryons that will be generically denoted by $\chi_i$) and the composite scalar (the dynamical axion $a$), 
\begin{equation}
j^\mu_{\text{PQ}}=f_{\text{PQ}}\partial^\mu a + \sum_{i}Q_{\chi_i}\,(\chi_i^\dagger\,\bar\sigma^\mu \,\chi_i)\,.
\end{equation}
This current  is classically conserved but it has a QCD anomaly,
\begin{equation}
\partial_{\mu}j^\mu_{\text{PQ}}=N\,\frac{\alpha_s}{8\pi}\, G\tilde{G}\,.
\end{equation}
This ward identity is reproduced by the following effective Lagrangian:  
\begin{equation}
\mathcal{L}_{\text{eff}}=\frac{1}{2} \partial^\mu a\partial_\mu a +\,\frac{\partial_\mu a}{f_{\text{PQ}}}\,\sum_{i} Q_{\chi_i}\,(\chi_i^\dagger\,\bar\sigma^\mu \,\chi_i)\,+\, N\,\frac{\alpha_s}{8\pi}\,\frac{a}{f_{\text{PQ}}} G\tilde{G}\,, 
\label{Eq:Eff Lag axion} 
\end{equation}
where the PQ symmetry is realized by the shift of the axion in Eq.\,(\ref{eq:shift}) with $\chi_i$'s kept invariant.

\subsubsection*{Relation between $\fpq$ and $\Lambda_5$ in Na\"ive Dimensional Analysis}
The effective Lagrangian obtained above 
can be rewritten in terms of a complex field satisfying $U\,U^\dagger=1$, 
\begin{equation}
U=e^{i\,5 a/f_{\text{PQ}}}\,,
\end{equation}
where the factor 5 is introduced to take into account that the physical domain of the axion field is $ a/f_{\text{PQ}}\in [0,2\pi/5)$, as shown in  App.~\ref{App:Axion domain}.
 The result is 
\begin{equation}
\mathcal{L}_{\text{eff}}=\frac{1}{2}
\left(\frac{f_{\text{PQ}}}{5}\right)^2\, \partial^\mu U^*\partial_\mu U - 5\,\frac{\partial_\mu a}{f_{\text{PQ}}}\,\left(\chi^\dagger\,\bar\sigma^\mu \,\chi\right) \,+\, \ldots \label{Eq:lagU axion and chi} 
\end{equation}
where the kinetic term is canonically normalized.  In this equation, the sum over composite baryons only shows explicitly the unique type of baryon made out of three fermions, which happens to be the baryon $\chi$  with PQ charge $Q_\chi=-5$  defined 
 in Eq.\,(\ref{onlybaryon}),  albeit now being massive.  

Applying Na\"ive Dimensional Analysis  (NDA)~\cite{Cohen97p301,Gavela16p485} to the Lagrangian in Eq.\,(\ref{Eq:lagU axion and chi}), it follows that 
\begin{equation}
\mathcal{L}_{\text{eff}}=\left(\frac{\Lambda_5}{4\pi }\right)^2\, \partial^\mu U^*\partial_\mu U +\,\left(\frac{4\pi}{\Lambda_5}\right)\,\partial_\mu a\,\left(\chi^\dagger\,\bar\sigma^\mu \,\chi\right) \,+\, \ldots
\label{Eq: NDA U and axion}
\end{equation}
leading to the identification 
\begin{equation}
\Lambda_5 \simeq \frac{4\pi \, \fpq}{5}\,.
\label{NDAsat}
\end{equation}
Customarily, the axion scale $f_a$ is defined reabsorbing in it the QCD anomaly factor $N$,
\begin{equation}
f_a\equiv \frac{f_{\text{PQ}}}{N}\,.
\end{equation}

\subsubsection*{Coupling to gluons and Domain Walls}
Because of the periodicity of the instanton potential, the anomalous coupling of the axion to gluons breaks explicitily $U(1)_{PQ}$ to a discrete symmetry $S(m)$,
\begin{equation}
S(m): a\longrightarrow a + \frac{2\pi m}{N} \fpq, \qquad m \in \mathbb{Z}\,.
\label{Eq: Discrete N}
\end{equation}
Nevertheless, not all  $S(m)$ transformations are nontrivial, as some of them are equivalent via gauge transformations (see App.~\ref{App:Axion domain}). The physical discrete symmetry corresponds to the quotient $S_{\rm phys}=S/{\mathbb Z}_5$, where ${\mathbb Z}_5$ is the center of the $SU(5)$ group~\cite{Ernst:2018bib}. 
This implies that the QCD potential has dim$[S_{\rm phys}]$ degenerate minima and therefore a number of domain walls $N_{DW}={\rm dim}[S_{\rm phys}]$ will be generated when the axion field takes a VEV, as this breaks spontaneously the discrete symmetry, 
\begin{equation}
N_{\rm DW}= \frac{|N|}{5}\,. 
\label{NDW}
\end{equation}
 Any theory with  $N_{\text{DW}}>1$ has a domain wall problem: domain walls could dominate the energy density of the universe and overclose it.  
It will be seen further below that in our theory indeed $N_{\text{DW}}>1$, and in consequence 
 a pre-inflationary PQ-transition will be assumed to avoid this issue (see e.g. \cite{Kawasaki:2013ae} and references therein).
Besides, we also assume that the highest temperature after inflation is lower than  $\Lambda_5$ to avoid the production
of massive particles in the dynamical sector, as some of them are stable due to the ${\mathbb Z}_2$ unbroken subgroup of the PQ symmetry, leading to an unacceptably large relic density.
 
\subsection{Planck suppressed operators}
It has been argued that quantum gravity may violate all global symmetries. In particular, Planck suppressed  operators which are not PQ invariant could be dangerous for axion solutions to the strong CP problem, since they can unacceptably displace the minimum of the axion potential from the CP conserving point.

Within our model, because of gauge invariance and chirality, the lowest dimensional  operator of this type has mass dimension nine, as previously argued:  it is the operator in Eq.\,(\ref{Eq:PQ condensate SU(n)}), whose VEV breaks PQ spontaneously.  This significantly strong Planck suppression suggest that our model can be  protected from those gravitational issues. This is to be contrasted with the usual expectation in axion models which allow lower dimension effective operators of gravitational origin, e.g. dimension five couplings as in Eq.\,(\ref{gravity-5}).  

The prefactors of the effective operator are relevant and they can be settled using  NDA~\cite{Cohen97p301,Gavela16p485}, resulting in:
\begin{equation}
\mathcal{L}_{\cancel{PQ}}=c\,\frac{1}{4\pi}\frac{1}{M^5_{\rm Pl}}\, \frac{1}{2!\,4!}\,\mathbf{\bar{5}\,\bar{5}\,10\,\bar{5}\,\bar{5}\,10}\,,
\label{QG}
\end{equation}
at around the  Planck scale.
Here, $c$ would be generically of order one and a combinatorial factor due to the presence of identical fields has been explicitly included in the definition of the operator.\footnote{Consistently, this would correspond to a combinatorial factor of 1 in the corresponding Feynman rules.}
In order to quantify its impact  on the location of the minimum of the axion potential, it is necessary to express it in terms of the low-energy composite fields.  NDA leads to 
\begin{equation}
\mathcal{L}_{\cancel{PQ}}=c\,\frac{(4\pi)^2}{2!\,4!}\left(\frac{N}{5}\right)^9\frac{f_a^9}{M^5_{\rm Pl}}e^{-i\frac{10}{N}a/f_a}+\text{h.c.}\,.
\label{effQG}
\end{equation}
The resulting axion potential,  including as well the QCD contribution reads~\footnote{The QCD axion potential is approximated here by a cosine dependence, 
since we are only interested in the displacement of the minimum where that approximation is perfectly valid. For the correct dependence using chiral Lagrangians 
at NLO see Ref.~\cite{diCortona:2015ldu}.} 
\begin{equation}
V (a)=-m_a^2 f_a^2\,\text{cos}\left(\frac{a}{f_a}\right) - 
 c\,\frac{(4\pi)^2}{4!}\left(\frac{N}{5}\right)^9\frac{f^9_a}{M^5_{\rm Pl}}\text{cos}\left(\frac{10}{N}\frac{a}{f_a}+ \delta\right)\,, 
\end{equation}
where $\delta$ is the relative phase between the Planck-suppressed operator in Eq.\,(\ref{effQG}) and the QCD 
vacuum parameter. The displacement of the axion VEV with respect to the CP conserving minimum is then given by
\begin{equation}
|\Delta \bar{\theta}_{eff}| = |c\,\text{sin}(\delta)|\,\frac{2\,(4\pi)^2}{4!}\left(\frac{N}{5}\right)^8\frac{f^7_a}{M^5_{\rm Pl}\,m_a^2}\,, 
\end{equation} 
which is strongly constrained by the experimental limit on the neutron electric dipole moment (EDM). 
For a given implementation of the $SU(5)$ theory, this indicates an upper bound on the $f_a$ value needed to avoid to fine-tune the coefficient of the gravitationally induced effective operator.

There is a certain degree of uncertainty when using power counting arguments in the present context, though, which may change the prefactors significantly. 
As illustration, if $f_{a}$ is taken as the PQ physics scale (instead of saturating it by $\Lambda_5 \sim 4\pi f_{PQ}/5 $ as in NDA), 
the operator in Eq.\,(\ref{QG}) would translate into
\begin{equation}
\mathcal{L}_{\cancel{ PQ}}=c\,\frac{1}{2!\,4!}\,\frac{f_a^9}{M^5_{\rm Pl}}e^{-i\frac{10}{N}a/f_a}+\text{h.c.}\,,
\label{Eq:Even more naive}
\end{equation}
instead of Eq.\,(\ref{effQG}). 
The displacement induced on the effective QCD vacuum angle would then be significantly smaller, depending on the value of the anomaly factor $N$ in a given realization of the chiral confining SU(5) theory. 

\vspace{0,5cm}

We will apply next the analysis above to two examples of the confining chiral $SU(5)$  theory, which differ in the QCD charges of the exotic  fermions $\{\psi_{\bar{5}}, \psi_{10}\}$, corresponding respectively to a reducible and irreducible QCD representation $\mathbf{R}$. 
In the first model $R=3+\bar{3}$, while $R=8$ will be assumed in the second model. 
While the former requires four exotic fermions (instead of just two for the second option), its matter content is smaller in terms of number of degrees of freedom. 
 
 \section{Model I:  color-triplet fermions} \label{Sec:3+3}
  In the first model, the exotic $\{\psi_{\bar{5}}, \psi_{10}\}$ fermions appear in the fundamental representation of QCD, alike to SM quarks, with
   \begin{equation}
   \mathbf{R}=\mathbf{3}+\mathbf{\bar{3}}\,,
   \end{equation}
    as shown in Table~\ref{II:tab:matterContent-triplet}. The $\left[SU(3)_c\right]^3$ anomalies are then automatically cancelled due to the the four distinct $SU(5)$ fermions present. 
    Being the latter massless, at the classical level this spectrum has four independent $U(1)$ global chiral symmetries. One combination is broken by non-perturbative $SU(5)$ effects, and three would remain unbroken for vanishing $\alpha_s$, one of them being  the PQ symmetry. 
The dimension of the (pseudo)real representation is then
  \begin{equation}
  n=6\, . 
  \end{equation}
  As indicated in Eq.~(\ref{chiralsym}), the global chiral symmetries correspond to $SU(6)_{\bar 5} \times SU(6)_{10}\times U(1)_{PQ}$ for $\alpha_s = 0$, which is explicitly broken by QCD down to
\begin{equation}
 SU(6)_{\bar{5}} \times SU(6)_{10}\times U(1)_{PQ} \xrightarrow{\alpha_s \neq 0}SU(3)_c\times U(1)_{V,\,\bar{5}} \times U(1)_{{V,\,10}}\,. 
 \label{TwoU1}
\end{equation}
That is, only QCD plus two global $U(1)$ symmetries remain unbroken for $\alpha_s \neq 0$, 
while $U(1)_{PQ}$ is broken by the non-perturbative QCD effects. 
The two surviving $U(1)$ symmetries 
are the left-over of the four classical $U(1)$ symmetries associated to the four independent massless fermions of this model (see Table~\ref{II:tab:matterContent-triplet}), as two were explicitly broken by anomalous couplings at the quantum level:  respectively $SU(5)$ and QCD interactions.  

\begin{table}[h!]
\begin{align*}
\begin{array}{c|c|c||c||c|c}
	& SU(5)	& SU(3)_c	& U(1)_{PQ} & U(1)_{V,\,\bar{5}} & U(1)_{V,\,10} \\
\hline
\psi_{({\bar{5}},3)}	    & \mathbf{\bar{5}}		& \mathbf{3}        &-3 & 1   & 0 \\
\psi_{(\bar{5},\bar{3})}	& \mathbf{\bar{5}}		& \mathbf{{\bar{3}}}&-3 & -1  & 0 \\
\psi_{(10,3)}	\,	        & \mathbf{10}		      & \mathbf{3}        &+1 & 0   & 1 \\
\psi_{(10,\bar{3})}\,     & \mathbf{10}		      & \mathbf{{\bar{3}}}&+1 & 0   & -1
\end{array}
\end{align*} 
\caption{Model I: charges of exotic fermions under the confining gauge group $SU(5)\times SU(3)_c$, the PQ symmetry and the spontaneously broken global $U(1)$ symmetries.  The left-handed Weyl fermions $\psi_{\bar{5}}$ and $\psi_{10}$ are massless and singlets of the SM electroweak gauge group; their QCD representation has been indicated as an additional  subscript.}
\label{II:tab:matterContent-triplet}
\end{table}
 
 The question of whether the QCD group $SU(3)_c$ is indeed the surviving unbroken group after chiral symmetry breaking, as indicated in Eqs.~(\ref {Eq:chiral condensate SU(n)}), (\ref{chiralbreaking}) and (\ref{totalbreaking}), deserves a specific discussion. 
To see this, let us note that an $SO(6)$ subgroup of the global symmetry $SU(6)_{\bar 5}\times SU(6)_{10}$ satisfies the 't Hooft anomaly consistency conditions. 
Besides, the condensates ${\langle \mathbf{10\,10\,10\,\bar{5}}\rangle}$ and $\langle\mathbf{\bar{5}\,\bar{5}\,10\,\bar{5}\,\bar{5}\,10}\rangle$ can be $SO(6)$ singlets.
This means  that the unbroken subgroup $G$ of the global symmetry $SU(6)_{\bar 5}\times SU(6)_{10}$ contains $SO(6)$, i.e.  $G \supset SO(6)$.%
\footnote{Our arguments do not depend on whether $G = SO(6)$ or $G \supsetneq SO(6)$, although we expect that $G = SO(6)$.}
The $SU(3)$ subgroup of $SO(6)$ is then obtained by identifying the vector representation of $SO(6)$ to be $\mathbf{3} + {\mathbf{\bar 3}}$.
Therefore, it is clear that an $SU(3)$ global symmetry remains unbroken below the confinement scale.

It should be noted  that an $SO(6)$ subgroup of $SU(6)_{\bar 5}\times SU(6)_{10}$ is not uniquely determined, and hence, the unbroken $SO(6)$ 
is not in general aligned to the one which contains $SU(3)_c$ for $\alpha_s = 0$.
However, it has been argued that, among the  possible condensate channels, the minimum of the potential corresponds to the one preserving 
QCD for $\alpha_s\ne 0$~\cite{Dobrescu:1996jp}. 
Thus, we find that it is most likely that the $SU(5)$ dynamics with the non-vanishing  chiral and PQ condensates in Eqs.~(\ref{Eq:chiral condensate SU(n)}) and (\ref{Eq:PQ condensate SU(n)}) preserves  $SU(3)_c$.

The $U(1)_{V\,,\bar 5}$ and $U(1)_{V\,,10}$ symmetries are generically broken by those condensates.
In fact, the chiral condensate in Eq.~\ref{Eq:chiral condensate SU(n)} breaks spontaneously $U(1)_{\bar 5}\times U(1)_{10}$ down to a $U(1)$,  
where the number of  positive and  negative charges with respect to this $U(1)$ is balanced at the QCD preserving vacuum.
The PQ condensate could also break this remaining $U(1)$ if the  quarks in the condensates
are all either in the ${\mathbf 3}$  or in the ${\mathbf {\bar 3}}$ representation of QCD. Accordingly,  the model predicts  one or two additional pGBs which obtain tiny masses from the 
higher dimensional gravitational operators in Eq.\,(\ref{QG}). 
As those pGBs decouple from the thermal bath at a temperature much higher than the weak scale,
the contribution of each pGB to the effective number of relativistic species is suppressed, i.e. ${\mit \Delta}N_{\rm eff} \simeq 0.03$,
and hence  the model is consistent with the current constraint  $N_{\rm eff} =2.99^{+0.34}_{-0.33}$~\cite{Aghanim:2018eyx}. 

In this model, the PQ current in Eq.\,(\ref{PQcurrent}) takes the form
\begin{align}
j^\mu_{\text{PQ}}=&-3 \psi_{(\bar{5},3)}^\dagger \bar \sigma ^\mu \psi_{(\bar{5},3)} - 3
\psi_{(\bar{5},3^*)}^\dagger \bar \sigma ^\mu \psi_{(\bar{5},3^*)} \\
&+ \psi_{(10,3)}^\dagger \bar \sigma ^\mu \psi_{(10,3)} + 
\psi_{(10,3^*)}^\dagger \bar \sigma ^\mu \psi_{(10,3^*)}=f_{\text{PQ}}\partial^\mu a\,.
\end{align}
For fermions in the fundamental of QCD ($T(\mathbf{\bar{3}})=T(\mathbf{3})=1/2$),   the QCD anomaly factor and the domain wall number in Eqs.~(\ref{Eq: PQ QCD anomaly}) and (\ref{NDW}) are then, respectively,
     \begin{equation}
N=  - 10\,,\qquad  N_{\rm DW}= 2\,.
\label{N3}
  \end{equation}
The resulting domain wall problem is avoided here by the assumption of pre-inflationary PQ transition, as earlier explained. 

\begin{figure}
\centering
\includegraphics[scale=.5]{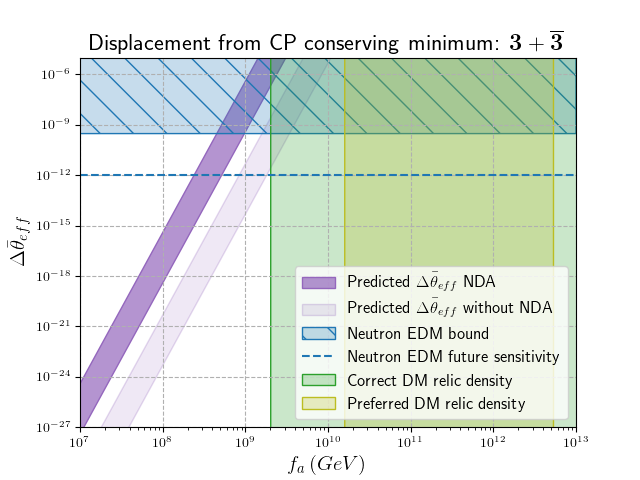}\includegraphics[scale=.5]{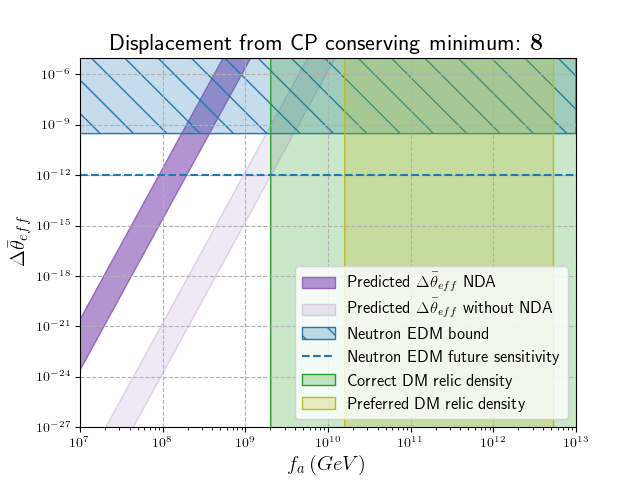}
\caption{Displacement of the CP conserving minimum due to the presence of the Planck suppressed operator for $|c\,\text{sin}(\delta)|  \in (0.001,1)$, assuming NDA. The regions excluded by the experimental limits on  the neutron EDM are depicted in blue, while  future prospects are indicated by a dashed blue line. The $f_a$ values that suffice to account for the full content of dark matter in the pre-inflationary scenario are depicted in green.}
\label{II:Fig:Displacement1}
\end{figure}

  \subsubsection*{Planck suppressed operators}
\begin{figure}
\centering
\includegraphics[scale=.5]{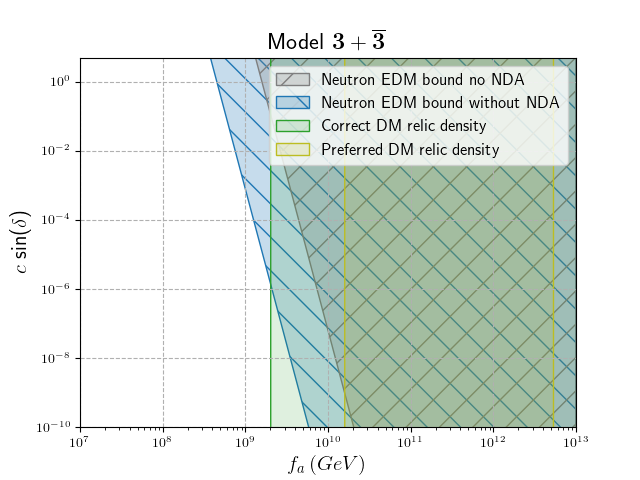}\includegraphics[scale=.5]{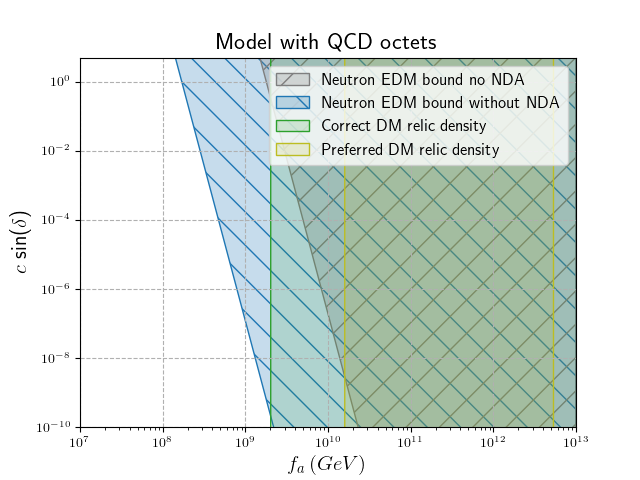}
\caption{Allowed values for the Planck suppressed operator coefficient $|c\,\text{sin}(\delta)|$, for axion dark matter compatible with neutron EDM bounds. }
\label{Fig:Displacement2}
\end{figure}

 For the value of $N$ in Eq.~(\ref{N3}),  the displacement induced on the QCD $\bar{\theta}$ parameter by  the NDA estimation of the Planck suppressed operator in Eq.\,(\ref{effQG}) is illustrated in Fig.\,\ref{II:Fig:Displacement1} (left panel). The figure  also depicts the stringent constraint imposed by the experimental bound on the neutron EDM~\cite{Baker:2006ts}, 
 which for the most conservative estimates~\cite{Engel:2013lsa} translates into the requirement
  \begin{align}
\mathbf{3}+ \bar{\mathbf{3}}\, \text{  Model: } \qquad f_a\lesssim 
(\,4.5\times10^8\,,\, 1\times10^9\,) \,{\rm GeV}\,,\qquad 
{\rm for} \, |c\,\text{sin}(\delta)|  \in (0.001,1)\,. 
\label{EDMtriplet}
\end{align}
 The softer constraint that follows if NDA is disregarded and  substituted by the estimation stemming from Eq.\,(\ref{Eq:Even more naive}) is  also depicted.\footnote{The explicit breaking can be further suppressed if, for example, we assume supersymmetry with R-symmetry.
In such cases, $f_a$ in the preferred value for the DM relic density is also allowed, though we do not pursue 
such possibilities further in this paper.} The degree of tuning of the  operator coefficient is illustrated in Fig.\,\ref{Fig:Displacement2} (left panel). 

\subsection*{ Axion dark matter}

In the misalignment mechanism, the relic axion abundance $\Omega_a$ depends then on two variables: the axion decay constant $f_a$, and the initial misalignment angle $\theta_{i}=a_i/f_a$. For $|\theta_{i}|\ll \pi$ it reads \cite{Saikawa17p083}
\begin{equation}
\Omega_a\,h^2=0.35\left(\frac{\theta_i}{0.001}\right)^2 \left(\frac{f_a}{3\times 10^{17}\text{GeV}}\right)^{1.17} \,,
\end{equation}
where $h$ is the present Hubble parameter.
 If  axions were to explain the total relic dark matter density 
 $\Omega_{\text{DM}}\,h^2\simeq 0.12$~\cite{Aghanim:2018eyx}, 
 the $f_a$ value required for
 an initial misalignment angle  in the range $\theta_{i}\in (0.1,3)$ is  
\begin{equation}
f_a\simeq 2\times 10^{10}-5\times 10^{12} \,\text{GeV}\,. 
\label{fadmpreferred}
\end{equation}
 However,  for values of $\theta_{i}\sim \pi$, the anharmonicities of the QCD potential are important and $f_a$ can be as low as~\cite{Visinelli:2009zm,Saikawa17p083,Tanabashi18p030001}
 \begin{equation}
 f_a\sim 2\times 10^{9}\,\text{GeV}\,.
 \label{dmfamin}
 \end{equation}
These two estimations of the $f_a$ values  that allow axions to constitute all the dark matter of the universe have been depicted in Fig.\,\ref{II:Fig:Displacement1} by green bands dubbed, respectively, ``preferred'' and ``correct'' densities.
 The lower  $f_a$ value in Eq.~(\ref{dmfamin})  is about a factor of two too large to be compatible with that required in Eq.\,(\ref{EDMtriplet}) by the neutron EDM bounds. This option requires a fine-tuning of the coefficient $c$ of the Planck suppressed operator of $\mathcal{O}(10^{-7})$, to be compared with the typical adjustment by 54 orders of magnitude in axion models with dimension five Planck-suppressed operators.  Furthermore, for a misalignment angle close to $\pi$ and low inflation scales, lower values of $f_a$ are possible and the fine-tuning of $c$ could be avoided altogether, even in this most conservative case of the NDA estimate of the effect.  Conversely, would the NDA prefactors be disregarded, $\mathcal{O}$(1) coefficients for the Planck suppressed operator are seen to be allowed in a large fraction of the parameter space.

  \section{Model II:  color-octet fermions}
  
   We consider here an alternative realization with only one $\{\psi_{\bar{5}}, \psi_{10}\}$ set of two fermions charged under $SU(5)$ and belonging to the adjoint representation of QCD, see Table~\ref{II:tab:matterContent-8}. All gauge anomalies cancel then automatically. This model is less economical than Model I, though, from the point of view of the total number of exotic degrees of freedom. \begin{table}[h!]
\begin{align*}
\begin{array}{c|c|c||c|c||c}
	& SU(5)	& SU(3)_c	& U(1)_{PQ}  \\
\hline
\psi_{\bar{5}}	& \mathbf{\bar{5}}		& \mathbf{8}	     &-3 \\
\psi_{10}	\,	& \mathbf{10}		& \mathbf{8}	   &+1  
\end{array}
\end{align*} 
\caption{Model II: charges of exotic fermions under the confining gauge group $SU(5)\times SU(3)_c$. Their PQ charges are shown as well.   The left-handed Weyl fermions $\psi_{\bar{5}}$ and $\psi_{10}$ are massless and singlets of the SM electroweak gauge group.}
\label{II:tab:matterContent-8}
\end{table}

In the limit of vanishing $\alpha_s$ the number of flavours of the $SU(5)$ fermionic sector is
\begin{equation}
  n=8\,, 
  \end{equation}
  and thus the global chiral symmetries of the $SU(5)$ Lagrangian correspond to 
\begin{equation}
 SU(8)_{\bar{5}} \times SU(8)_{10}\times U(1)_{PQ} \xrightarrow{\alpha_s \neq 0}SU(3)_c\,. 
\end{equation}
In consequence, in this model only QCD remains unbroken for $\alpha_s \neq 0$, and hence  no light pNGs appear associated with the spontaneous breaking 
of the non-abelian global symmetries.

To see whether the QCD gauge group remains ultimately unbroken,   note that an $SO(8)$ subgroup of the global symmetry $SU(8)_{\bar 5}\times SU(8)_{10}$ 
satisfies the 't Hooft anomaly consistency conditions,  while the condensates ${\langle \mathbf{10\,10\,10\,\bar{5}}\rangle}$ and $\langle\mathbf{\bar{5}\,\bar{5}\,10\,\bar{5}\,\bar{5}\,10}\rangle$ 
can be $SO(8)$ singlets.
In this case, we find that the  unbroken subgroup $G$ contains $SO(8)$, i.e.  $G \supset SO(8)$.
The $SU(3)$ subgroup of $SO(8)$ is realized as the special maximal embedding where the vector representation of $SO(8)$ is identified 
with the octet of $SU(3)$ (see e.g.~\cite{Ramond:2010zz}).
Thus, it is again clear that an $SU(3)$ global symmetry remains unbroken below the confinement scale, with non-vanishing  ${\langle \mathbf{10\,10\,10\,\bar{5}}\rangle}$ and $\langle\mathbf{\bar{5}\,\bar{5}\,10\,\bar{5}\,\bar{5}\,10}\rangle$ condensates.
Finally, the $SO(8)$ symmetry is aligned with that containing $SU(3)_c$ once $\alpha_s \neq 0$ is taken into account.
This shows that, also in this model, it is most likely for the $SU(5)$ dynamics to preserve $SU(3)_c$. 

For fermions in the adjoint of QCD ($T(\mathbf{R})=3$),   the QCD anomaly factor and the domain wall number in Eqs.\,(\ref{Eq: PQ QCD anomaly}) and (\ref{NDW}) are, respectively,
   \begin{equation}
N=  - 30\,,\qquad N_{\rm DW}= 6\,.
\label{Noctet}
  \end{equation}

   \subsubsection*{Planck suppressed operators}
Fig.\,\ref{II:Fig:Displacement1} (right panel)  shows the displacement induced  by  the  operator in Eq.\,(\ref{effQG}) on the QCD vacuum  parameter, for the value of $N$ expected from NDA, see Eq.\,(\ref{Noctet}), which implies 
 the requirement
  \begin{align}
\mathbf{8}\, \text{  Model: } \qquad f_a\lesssim 
(\,1.7\times10^8\,,\, 3.7\times10^8\,) \,{\rm GeV}\,,\qquad 
{\rm for} \, |c\,\text{sin}(\delta)|  \in (0.001,1)\,,
\label{EDMoctet}
\end{align}
to comply with the experimental bound on the neutron EDM. This constraint is stronger than that for Model I for QCD-triplet exotic fermions, Eq.\,(\ref{EDMtriplet}).  A softer constraint if NDA was disregarded in the estimation is also illustrated.

\subsection*{ Axion dark matter}
 The comparison between Eq.\,(\ref{EDMoctet}) and the $f_a$ ranges in Eqs.\,(\ref{fadmpreferred}) and (\ref{dmfamin}) shows that this model with exotic fermions in the adjoint of QCD is more in tension than model I,   if axions are to explain all the dark matter of the universe without recurring to fine tunings.   Fig.\,\ref{II:Fig:Displacement1} (right panel) illustrates this situation. For the NDA estimation of Planck suppressed couplings,  $f_a$ as required by dark matter is a factor of  five   too large with respect to the neutron EDM constraint; this translates into the requirement of a $\mathcal{O}(10^{-10})$ fine-tuning of the coefficient $c$ of the Planck suppressed operator,  
as illustrated in Fig.\,\ref{Fig:Displacement2} (right panel).    Alternatively, the present model could explain a subdominant fraction of the dark matter content. 
 
 A comparison without NDA power counting estimates is also illustrated: 
  non-fine tuned values of the coefficient $c$ are then compatible with the axion accounting   for the ensemble of dark matter, while  complying with EDM limits.   Overall, the uncertainty on  the estimations of non-perturbative gravitational effects, and on the $f_a$ values required to account for dark matter, is large enough to still consider 
 this model as a candidate scenario for purely axionic dark matter.


\newpage
\section{Conclusions}

We have presented a novel composite axion theory that solves the strong CP problem and has as singular features:

\begin{itemize}
\item A  gauge confining symmetry which is chiral, unlike usual axicolor models which use vectorial fermions.  In consequence, the PQ symmetry is automatic, without any need to invoke extra symmetries.

\item Exotic $SU(5)$ fermions in (pseudo)real representations of QCD.

\item Inherent protection from dangerous quantum non-perturbative gravitational effects. 

\end{itemize}
The gauge group selected and illustrated here is chiral $SU(5)$ with two massless fermions in its $\mathbf{\bar{5}}$ and $\mathbf{10}$ representations and a confining scale much higher than that of QCD.  A new spectrum of composite bound states is expected.

We showed that  the `t Hooft anomaly conditions for the global symmetries of the exotic fermionic sector imply  that  the non-abelian global symmetries must  be spontaneously broken. The global abelian symmetries, e.g. the PQ symmetry, must also be spontaneously broken  for the theory to be phenomenologically viable, resulting in a dynamical invisible axion.
Furthermore, the PQ invariance is  the analogous of the $B-L$ symmetry in $SU(5)$ Grand Unified Theory (GUT).

We have determined the fermionic  operators with lowest dimension  which may condense and induce spontaneous breaking. Because of $SU(5)$ gauge invariance, six is the minimal dimension for the operator whose VEV may break  the exotic flavour symmetries. 
 An even higher dimensional condensate is needed  in order to break PQ invariance: the VEV of a dimension nine  operator. 
 The latter is also the lowest dimensional effective operator which could result from  gravitational quantum contributions, breaking explicitly the PQ symmetry, as these effects must respect gauge invariance. Its high dimensionality 
 is at the heart of the inherent protection of this theory with respect to the gravitational issue.

We have developed two complete ultraviolet completions of the chiral confining $SU(5)$ theory,   which only differ in the (pseudo)real QCD representations chosen for the exotic fermions:  a reducible  $\mathbf{3}+\mathbf{\bar{3}}$ representation for Model I, and the irreducible adjoint in model II.  The former is more economical in terms of the total number of degrees of freedom.  Both models are phenomenological viable and largely protected from quantum gravitational concerns.   Remarkably, in the case of exotic fermions in the fundamental of QCD, the $f_a$ range allowed  if axions are to explain the full dark matter content of the universe can be compatible with that required to avoid a fine-tuned coefficient for the Planck suppressed operator. For octet-colour fermions the compatibility is marginal but still possible. 

The basic novel idea of the construction is to use a chiral confining group, which provides an automatic implementation of  PQ invariance. 
The most economic avenue is to implement it via just two  exotic fermions in (pseudo)real representations of QCD. In this perspective, we have briefly explored other  confining groups as well. For instance, 
a chiral and confining gauge $SU(4)$ symmetry would be a viable alternative, although  it does not enjoy a sufficient protection from gravitational issues, at least in the case of only two exotic fermions.  Even the smaller  chiral confining $SU(3)$ symmetry is possible, although the versions with only  two exotic fermions require very high-dimensional representations of the confining group and, again, they are less protected from gravitational issues  than the $SU(5)$ case (see App.~\ref{App:SU(4)}).  Nevertheless, as the estimation of gravitational effects is somehow uncertain, it may be pertinent to  dedicate specific studies to these alternative directions.

\section*{Acknowledgments}

We acknowledge very interesting conversations and comments from Valery Rubakov, Mary K. Gaillard, Rachel Houtz, A. Manohar, Rocio del Rey and G. Villadoro.
M.B.G and P. Q. acknowledge IPMU at Tokyo University, where  this work was done. T.~T.~Y. is grateful to Sergei Kuzenko for the hospitality during his stay at The University of Western Australia.
This project has received support from the European Union's Horizon 2020 research and innovation programme under the Marie Sklodowska-Curie grant agreements No 690575  (RISE InvisiblesPlus) and  No 674896 (ITN ELUSIVES). M.B.G and P. Q. also acknowledge support from the 
 the Spanish Research Agency (Agencia Estatal de Investigaci\'on) through the grant IFT Centro de Excelencia Severo Ochoa SEV-2016-0597, as well as  from the ``Spanish Agencia Estatal de Investigaci\'on'' (AEI) and the EU ``Fondo Europeo de Desarrollo Regional'' (FEDER) through the project FPA2016-78645-P. 
The work of P.Q. was supported through a ``La Caixa-Severo Ochoa'' predoctoral grant of Fundaci\'on La Caixa.
This research was also supported in part by WPI Research Center Initiative, MEXT, Japan (MI, MY and TTY), and in part by JSPS Grant-in-Aid for Scientific Research No. 15H05889, No. 16H03991, No. 18H05542 (MI), No. 26104001, No. 26104009, No. 16H02176 (TTY), and No. 17H02878 (MI and TTY). TTY is a Hamamatsu Professor at Kavli IPMU.

\appendix
\newpage
\section{Alternative confining groups: $SU(3)$ and $SU(4)$} \label{App:SU(4)}

\subsection*{SU(4) Model}
It is also possible to construct a chiral axicolor model that fulfills the requirements explained in the introduction (see Sec.~\ref{introduction}) with an $SU(4)$ gauge group. 

\begin{table}[h!]
\begin{align*}
\begin{array}{c|c|c||c}
  & SU(4) & SU(3)_c&  U(1)_{PQ} \\
\hline
\psi_{\bar{4}}  & \mathbf{\bar{4}}   & \mathbf{8} & -3\\
\psi_{10}\, & \mathbf{10}    & \mathbf{1} & 4
\end{array}
\end{align*} 
\caption{Charges of exotic fermions under the confining gauge group $SU(4)\times SU(3)_c$.  The left-handed Weyl fermions $\psi_{\bar{4}}$ and $\psi_{10}$ are massless and singlets of the SM electroweak gauge group. }
\label{tab:matterContentSU(4)}
\end{table}

It is easy to check that this theory is free from gauge anomalies~\footnote{ $\left[SU(4)\right]^3$ anomaly: $8 A(\mathbf{\bar{4}})+A(\mathbf{10})=0$, since $ A(\mathbf{\bar{4}})=-1,\,\text{and }A(\mathbf{10})=8$.} and that the global $U(1)_{PQ}$ in Table~\ref{tab:matterContentSU(4)} is exact at the classical level but explicitly broken by $SU(3)_c$ instantons, solving therefore the strong CP problem \`a la Peccei-Quinn.

However we will not study this model further since it lacks special protection against PQ-violating gravity operators. Indeed the lowest dimensional non-renormalizable operators that break PQ and could be generated by quantum gravity effects are

\begin{equation}
\mathcal{L}_{Planck}\propto \frac{c}{M^2_{\rm Pl}}\, \frac{1}{4!}\,\mathbf{\bar{4}\, \bar{4}\, \bar{4}\, \bar{4}}\, + \frac{c}{M^2_{\rm Pl}}\, \frac{1}{4!}\,\mathbf{10}\, \mathbf{10}\, \mathbf{10}\, \mathbf{10}\,,
\label{QGSU4}
\end{equation}
and would lead to unacceptable deviations from the CP-conserving minimum (barring a fine-tuning of  $c$
 by several tens of orders of magnitude) and thus spoil the solution of the strong CP problem.

  \subsubsection*{Alternative SU(4)}
  It is possible to implement the confining gauge $SU(4)$ solution in a setup in which two exotic fermions belong to the adjoint of QCD, by considering higher $SU(4)$ representations, e.g. $\mathbf{\bar{35}}$ and $\mathbf{70}$ since 
\begin{table}[h!]
\begin{align*}
\begin{array}{c|c|c||c}
  & SU(4) & SU(3)_c&  U(1)_{PQ} \\
\hline
\psi_{\bar{35}}  & \mathbf{\bar{35}}   & \mathbf{8} & -98\\
\psi_{70}\,      & \mathbf{70}         & \mathbf{8} & 56
\end{array}
\end{align*} 
\caption{Charges of exotic fermions under the confining gauge group $SU(4)\times SU(3)_c$.  The left-handed Weyl fermions $\psi_{\bar{35}}$ and $\psi_{70}$ are massless and singlets of the SM electroweak gauge group. }
\label{tab:matterContentSU(4)}
\end{table}
$A(\mathbf{\bar{35}})=-112\,,\quad A(\mathbf{70})=+112 $, see Table~\ref{tab:matterContentSU(4)}.

\subsection*{SU(3) Model}
The idea of using a  chiral confining theory  as solution to the strong CP problem  can also be implemented with a confining $SU(3)$ gauge group, for instance via the fermionic content in Table~\ref{cantejondo}. 
\begin{table}[h!]
\begin{align*}
\begin{array}{c|c|c||c}
  & SU(3) & SU(3)_c&  U(1)_{PQ} \\
\hline
\psi_{\bar{15'}}  & \mathbf{\bar{15'}}   & \mathbf{R} & -119\\
\psi_{42}\, & \mathbf{42}    & \mathbf{R} & 35
\end{array}
\end{align*} 
\caption{Charges of exotic fermions under the confining gauge group $SU(3)\times SU(3)_c$.  The left-handed Weyl fermions $\psi_{\bar{15'}}$ and $\psi_{42}$ are massless and singlets of the SM electroweak gauge group. }
\label{cantejondo}
\end{table}

This theory is anomaly free since $A(\mathbf{\bar{15'}})=-A(\mathbf{42})=77$ and again the exotic fermions transform as pseudoreal representations $\mathbf{R}$ of the QCD group. However, the theory is not as protected against PQ breaking gravitational effect as the $SU(5)$ case, since the corresponding effective operators can appear at dimension six,
 \begin{equation}
\mathcal{L}_{Planck}\propto \frac{c}{M^2_{\rm Pl}}\, \frac{1}{2!2!}\,\mathbf{\bar{15'}\, \bar{15'}\, 42\, 42}\, \,,
\label{QGSU4}
\end{equation}
 and in consequence we will not further elaborate on this model.

\newpage

\section{Anomaly factors}
\label{App:Anomaly}
In this appendix we review the group theoretical factors that are relevant when computing the global or gauge anomalies in our theory. Let us consider a given conserved current $j^a_ \mu$ that corresponds to the symmetry associated to the generator $T^a$. In the presence of the gauge field $F_b$ the divergence of the current reads,
\begin{equation}
\partial^\mu j^a_ \mu=\frac{\alpha_i}{8\pi}C^{abc}_{group} F_{b\,\mu\nu} \tilde F_c^{\mu\nu}\,,
 \end{equation}
where $\tilde F^{\mu\nu}=\frac{1}{2}\epsilon^{\mu\nu\sigma\rho}F_{\sigma\rho}$,  the fine structure constant of the corresponding gauge interaction is denoted by  $\alpha_i=\frac{g_i^2}{4\pi}$ and the group theoretical factor $C_{group}$ is given by
\begin{equation}
C^{abc}_{group}=\sum Tr\left[T^a\{t_R^b,t_R^c\}\right]\label{general},
 \end{equation}
where the sum runs over all fermionic representations $\mathbf{R}$ of the gauge group $t_R^a$. Thoughout the paper the fermionic degrees of freedom will be expressed in terms of left-handed Weyl fermions.

This formula is used for three different cases,  depending on whether the groups are abelian or non-abelian and whether the anomaly is cubic or mixed.
\begin{itemize}
\item Non-abelian cubic anomalies:
\begin{equation}
\left[SU(N)\right]^3:\qquad C^{abc}_{group}=\sum_R Tr\left[t_R^a\{t_R^b,t_R^c\}\right]\equiv d^{abc}\sum _R A(\mathbf{R})\,,
\end{equation}
where $A(\mathbf{R})$ denotes anomaly coefficient or triality of the representation $\mathbf{R}$.
\item Abelian cubic anomalies:
\begin{equation}
\left[U(1)\right]^3:\qquad C_{group}=\sum_R Tr\left[Q_R\{Q_R,Q_R\}\right]=2\,\sum _R Q_R^3\,,
\end{equation}
where $Q_R$ denotes the $U(1)$ charge   of the corresponding fermion.
\item Mixed anomalies:
\begin{equation}
\left[SU(N)\right]^2\times U(1):\qquad C^{bc}_{group}=\sum_R Tr\left[Q_R\{t_R^b,t_R^c\}\right]\equiv \delta^{bc}\sum _R Q_R\, 2T(\mathbf{R})\,,
\end{equation}
where $T(\mathbf{R})$ is the Dynkin index of the representation $\mathbf{R}$. 
\end{itemize}
These group theoretical factors are tabulated \cite{Slansky:1981yr} and can also be computed with the Mathematica package LieART \cite{Feger:2012bs}.

\newpage

\section{Axion field domain}
 \label{App:Axion domain}
Our definition of the PQ symmetry according to the charges in Tab.~\ref{tab:matterContent}  corresponds to the following transformations:
\begin{align}
\psi_{10}& \longrightarrow e^{i\,\alpha}\,\psi_{10}\,, \nn \\
\psi_{\bar{5}}& \longrightarrow e^{-3\,i\,\alpha}\,\psi_{\bar{5}}\,,
\end{align}
where $\alpha$ is the rotation angle. However, the domain of $\alpha$ does not correspond to the full range $[0,2\pi)$ since some of these rotations are equivalent due to gauge transformations. In particular, the center of $SU(5)$ is the discrete symmetry $Z\left[SU(5)\right]=\mathbb{Z}_5$, that corresponds to the following gauge transformations:
\begin{align}
\psi_{10}& \longrightarrow e^{2\pi\,i\,k/5}\,\psi_{10}\,e^{2\pi\,i\,k/5}=e^{4\pi\,i\,k/5}\psi_{10} \,,\nn  \\
\psi_{\bar{5}}& \longrightarrow e^{-2\pi\,i\,k/5}\,\psi_{\bar{5}}\,,
\end{align}
for $k=\{0,1,2,3,4\}$. It is easy to see now that a PQ transformation with angle $\alpha=2\pi/5$ is gauge equivalent to $\alpha=2\pi$ with $k=2$.
As a consequence, the axion transforms under PQ as
\begin{equation}
\frac{a}{f_{\text{PQ}}}\longrightarrow \frac{a}{f_{\text{PQ}}} + \alpha \,     
\end{equation}  
 and its physical domain is 
 \begin{equation}
 \frac{a}{f_{\text{PQ}}}\in [0,2\pi/5)\,.
 \end{equation}

\newpage
\section{`t Hooft anomaly matching conditions: is  $SU(8)_{\bar{5}} \times SU(8)_{10}\times U(1)_{PQ}$ spontaneously broken?} 
\label{App:tHooft}
If the $SU(5)$ group confines without breaking the chiral symmetries in Table~\ref{tab:GlobalSymtHooft}, the consistency of the theory implies the existence of massless baryons in the low energy  that match the global anomalies of the high-energy theory. However, for some theories these `t Hooft anomaly matching conditions cannot be satisfied as a consequence of the properties of the fermionic representations.  It must be then concluded that these theories can only be realized via spontaneous breaking of its chiral symmetries. This will be the case for the $SU(8)_{\bar{5}} \times SU(8)_{10}$ chiral symmetry of our $SU(5)$ model.
\begin{table}[h!]
\begin{align*}
\begin{array}{c|c|c|c}
  & SU(8)_{\bar{5}} & SU(8)_{10}  &  U(1)_{PQ}    \\
\hline
\psi_{\bar{5}}  & \Box  & \mathbf{1}    & -3    \\
\psi_{10}\,     & \mathbf{1}     & \Box & 1      
\end{array}
\end{align*} 
\caption{Global chiral properties at the classical level, in the limit of vanishing $\alpha_s$. }
\label{tab:GlobalSymtHooft}
\end{table}

Let us first compute the global anomalies in the high energy theory (in terms of the fundamental quarks $\psi_{\bar{5}}$ and $\psi_{10}$):
\begin{align}
\left[SU(8)_{\bar{5}}\right]^3:& \qquad 5\times A(\Box)=5\,,\\
\left[SU(8)_{10}\right]^3:& \qquad 10\times A(\Box)=10\,,\\
 U(1)_{PQ}\times\left[SU(8)_{\bar{5}}\right]^2:& \qquad 5\times 2\,T(\Box)  Q_{\bar{5}}=-15\,,\\
 U(1)_{PQ}\times\left[SU(8)_{10}\right]^2 :& \qquad 10\times 2\,T(\Box)  Q_{10}=10\,,\\
\left[U(1)_{PQ}\right]^3:& \qquad 8 \left(5\,(Q_{\bar{5}})^3+10\,(Q_{10})^ 3\right)=-1000\,.
\label{Eq: global anomalies}
\end{align}

If chiral symmetries remain unbroken these anomalies will match those in the low energy theory in terms of the bound states. The simplest $SU(5)$ singlet that can be formed in this theory consists of three fundamental quarks, $\chi \equiv10\, \mathbf{\bar{5}} \, \mathbf{\bar{5}}\,$. Can it match the previous anomalies? The transformation properties of $\chi $ under the global symmetries are 
\begin{align}
SU(8)_{\bar{5}}:& \qquad \mathbf{8}\times \mathbf{8}= \mathbf{28} + \mathbf{36}\,,\\
SU(8)_{10}:& \qquad \mathbf{8}\,,\\
 U(1)_{PQ}:& \qquad  -3-3+1=-5\,.
\label{Eq: Transf chi global anomalies}
\end{align}
In consequence, there are two posible representations for the baryon $\chi$ under $SU(8)_{\bar{5}} \times SU(8)_{10}\times U(1)_{PQ}$:  $(\mathbf{28},\mathbf{8},-5)$ and $(\mathbf{36},\mathbf{8},-5)$. If the low energy contains a number $n_{28}$ and $n_{36}$ of baryons transforming under each representation respectively, then the anomalies are given by 
\begin{align}
\left[SU(8)_{\bar{5}}\right]^3:& \qquad 8\left( n_{28}\,A(\mathbf{28})+n_{36}\,A(\mathbf{36})\right)=32(n_{28} + 3n_{36})\,,\label{D9}\\
\left[SU(8)_{10}\right]^3:& \qquad  28\,n_{28}\,A(\mathbf{8})+36\,n_{36}\,A(\mathbf{8})=4(7\,n_{28} + 9\,n_{36})\,,\\
 U(1)_{PQ}\times\left[SU(8)_{\bar{5}}\right]^2:& \qquad 8\,Q_{\chi}\left( n_{28}\,2\,T(\mathbf{28})+n_{36}\,2\,T(\mathbf{36})\right)=-80(3\,n_{28} + 5n_{36})\,,\\
 U(1)_{PQ}\times\left[SU(8)_{10}\right]^2 :& \qquad Q_{\chi}\left( 28\,n_{28}\,2\,T(\mathbf{8})+36\,n_{36}\,2\,T(\mathbf{8})\right)=-20(7n_{28} + 9n_{36})\,,\\
\left[U(1)_{PQ}\right]^3:& \qquad 8 \left(28n_{28} + 36n_{36}\right)\,\left(Q_{\chi}\right)^3= -4000\left(7 n_{28} + 9 n_{36}\right).
\label{Eq: global anomalies}
\end{align}
It is easy to see that there is no way of matching these anomalies with $n_{28},\, n_{36} \in \mathbb{N}$. 
If we would alternatively consider 5-quark bound states, there are two options: $\mathbf{\bar{5}\,\bar{5}\,\bar{5}\,\bar{5}\,\bar{5}}$ and $\mathbf{10\,10\,10\,10\,10}$.
 
For the first one, $\mathbf{\bar{5}\,\bar{5}\,\bar{5}\,\bar{5}\,\bar{5}}$, the transformation properties are: 
\begin{align}
SU(8)_{\bar{5}}:& \qquad \mathbf{8}\times \mathbf{8}\times \mathbf{8}\times \mathbf{8}\times \mathbf{8}=& (\mathbf{\overline{56}})+4\ (\mathbf{\overline{504}})+(\mathbf{\overline{792}})+5\ (\mathbf{\overline{1008}})+6\
   (\mathbf{\overline{1512}}')  \\
   &&+5\ (\mathbf{\overline{1680}})+4\ (\mathbf{\overline{1848}})\,,\\
SU(8)_{10}:& \qquad \mathbf{1}\,,\\
 U(1)_{PQ}:& \qquad  5\,(-3)=-15\,.
\label{Eq: Transf chi global anomalies}
\end{align}
For $\mathbf{10\,10\,10\,10\,10}$ the transformation properties are: 
\begin{align}
SU(8)_{\bar{5}}:& \qquad \mathbf{1}\,,\\
SU(8)_{10}:& \qquad \mathbf{8}\times \mathbf{8}\times \mathbf{8}\times \mathbf{8}\times \mathbf{8}=& (\mathbf{\overline{56}})+4\ (\mathbf{\overline{504}})+(\mathbf{\overline{792}})+5\ (\mathbf{\overline{1008}})+6\
   (\mathbf{\overline{1512}}')  \\
   &&+5\ (\mathbf{\overline{1680}})+4\ (\mathbf{\overline{1848}})\,,\\
 U(1)_{PQ}:& \qquad  5\,(+1)=+5\,.
\label{Eq: Transf chi global anomalies}
\end{align}
Repeating the analogous exercise to that in Eqs.~(\ref{D9})-(\ref{Eq: global anomalies}), and using the properties of the representations of the 5-quark bound states in Table~\ref{tab:group coefficients}, it follows the same conclusion as before: the chiral symmetry must necessarily be spontaneously broken due to the impossibility of satisfying `t Hooft anomaly matching conditions.
\begin{table}[h!]
\begin{align*}
\begin{array}{c|c|c}
  R\mathbf{}& 2\,T(\mathbf{R}) & A(\mathbf{R}) \\
\hline
\overline{\mathbf{56} }  & 13 & -5\\
\overline{\mathbf{504}}  & 213 & 75\\
\overline{\mathbf{792}}  & 713 & 1287 \\
\overline{\mathbf{1008}} & 524 & 294\\
\overline{\mathbf{1680}} & 1088 & 1066\\
\overline{\mathbf{1512}}'& 883 & 777
\end{array}
\end{align*} 
\caption{Dynkin index $T(\mathbf{R})$ and anomaly factor $A(\mathbf{R})$ of the different representations of $SU(8)_{\bar{5}}$ that are contained in $\left[\mathbf{8}\right]^5$.}
\label{tab:group coefficients}
\end{table}

This does not mean, however, that the full $SU(8)_{\bar{5}} \times SU(8)_{10}\times U(1)_{PQ}$ is completely spontaneously broken. Some subgroup can remain unbroken. In particular, it is shown in the body of the paper that it is possible to leave unbroken the $U(1)_{PQ}$ with the baryon in Eq.~(\ref{onlybaryon}) satisfying the anomaly matching conditions. Nevertheless, this possibility is phenomenologically excluded due to the absence of coloured massless quarks in Nature. 


\newpage
\bibliographystyle{utphys.bst}

\providecommand{\href}[2]{#2}\begingroup\raggedright\endgroup
\end{document}